\documentclass[aps,reprint,floatfix,superscriptaddress]{revtex4-2}
\usepackage{amsmath,amssymb,amsthm}
\usepackage{physics}
\usepackage{graphicx}
\usepackage[bookmarks=false,colorlinks=true,linkcolor=blue,filecolor=blue,citecolor=blue,urlcolor=blue]{hyperref}
\usepackage[dvipsnames]{xcolor}
\usepackage{bm,ulem}

\begin{document}

\title{Topology and Spectrum in Measurement-Induced Phase Transitions}

\author{Hisanori Oshima}
\affiliation{Department of Applied Physics, University of Tokyo, Tokyo 113-8656, Japan}
\affiliation{Nonequilibrium Quantum Statistical Mechanics RIKEN Hakubi Research Team, RIKEN Pioneering Research Institute (PRI), 2-1 Hirosawa, Wako, Saitama 351-0198, Japan}

\author{Ken Mochizuki}
\affiliation{Department of Applied Physics, University of Tokyo, Tokyo 113-8656, Japan}
\affiliation{Nonequilibrium Quantum Statistical Mechanics RIKEN Hakubi Research Team, RIKEN Pioneering Research Institute (PRI), 2-1 Hirosawa, Wako, Saitama 351-0198, Japan}

\author{Ryusuke Hamazaki}
\affiliation{Nonequilibrium Quantum Statistical Mechanics RIKEN Hakubi Research Team, RIKEN Pioneering Research Institute (PRI), 2-1 Hirosawa, Wako, Saitama 351-0198, Japan}
\affiliation{RIKEN Center for Interdisciplinary Theoretical and
Mathematical Sciences (iTHEMS), RIKEN, Wako 351-0198, Japan
}

\author{Yohei Fuji}
\affiliation{Department of Applied Physics, University of Tokyo, Tokyo 113-8656, Japan}

\date{\today}

\begin{abstract}
Competition among repetitive measurements of noncommuting observables and unitary dynamics can give rise to a wide variety of entanglement phases.
Here, we propose a general framework based on Lyapunov analysis to characterize topological properties in monitored quantum systems through their spectrum and many-body topological invariants.
We illustrate this framework by analyzing (1+1)-dimensional monitored circuits for Majorana fermions, which are known to exhibit topological and trivial area-law entangled phases as well as a critical phase with sub-volume-law entanglement.
Through the Lyapunov analysis, we identify the presence (absence) of edge-localized zero modes inside the bulk gap in the topological (trivial) phase and a bulk gapless spectrum in the critical phase.
Furthermore, by suitably exploiting the fermion parity with twisted measurement outcomes at the boundary, we construct a topological invariant that distinguishes the two area-law phases and dynamically characterizes the critical phase.
Our framework thus provides a general route to extend the notion of bulk–edge correspondence to monitored quantum dynamics.
\end{abstract}

\maketitle

\paragraph*{Introduction--}
\label{sec:Introduction}
Topology is undoubtedly one of the most essential concepts in understanding stable phases of matter \cite{Wen17}.
For the ground states of local Hamiltonians with a finite excitation gap, states belonging to distinct topological phases cannot be smoothly deformed to each other without closing the gap, owing to discrete characters of their topological invariants \cite{Chen10}.
In particular, for symmetry-protected topological phases \cite{Gu09, Pollmann10, Chen11, Fidkowski11, Lu12, Chen13}, the nontrivial topology of the bulk leads to gapless boundary states robust against any symmetry-preserving perturbations.
This phenomenon, known as the bulk-edge correspondence, has remarkable consequences in the physical properties of the topological phases and has been intensively studied in the past few decades \cite{Hasan10, Qi11, Senthil15, Witten16, Chiu16}.

Despite notable successes in equilibrium systems, topology in out-of-equilibrium systems caused by the external environment has yet to be fully understood. 
In particular, it is still elusive how the topology plays a role in monitored quantum systems, which have garnered significant interest recently as they host novel nonequilibrium phases concerning, e.g., entanglement \cite{Li18, Skinner19, Chan19, Szyniszewski19, Li19, Jian20, Bao20, Choi20, Gullans20PRX, Gullans20PRL, Potter22, Fisher23}.
Indeed, repetitive measurements stabilizing distinct topological phases can lead to phase transitions between different entanglement phases, which can be probed by topological entanglement entropy or purification dynamics \cite{Lavasani21, Lavasani21PRL, Klocke22, Lavasani23, Sriam23, Zhu24, Morral-Yepes23, Kuno23, Nehra24, Kelson-Packer24, Klocke24, Nahum20, Bao21, Jian23, Klocke23, Kells23, Pan24} (see also \cite{Lang20, Sang21, Ippoliti21}).
However, intrinsic spatiotemporal randomness caused by the probabilistic nature of measurement outcomes makes it difficult to generalize more standard notions of topological phases, such as bulk topological invariants and gapless edge modes, to monitored systems.

\begin{figure}
\includegraphics[width=0.47\textwidth]{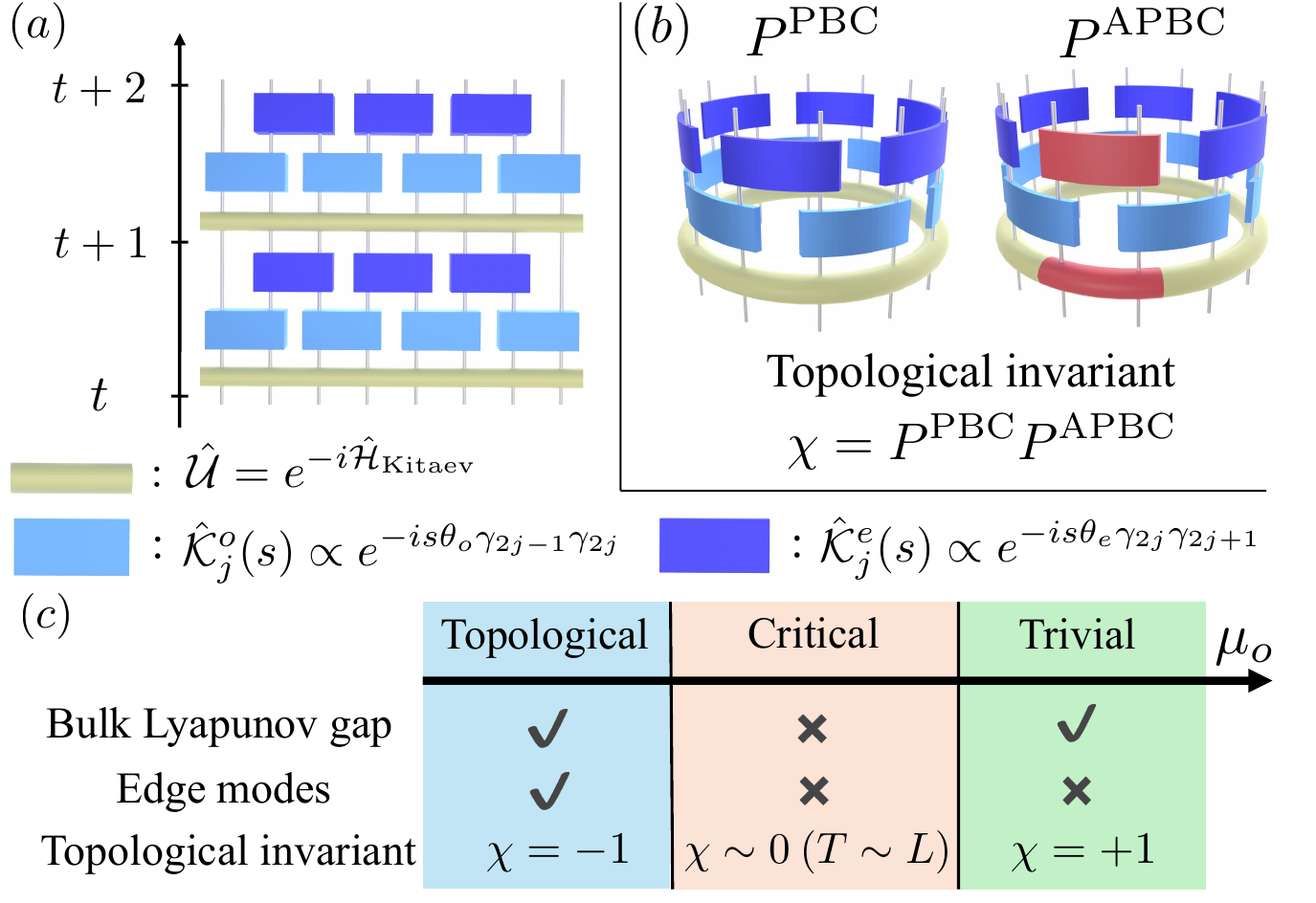}
\caption{
(a) Monitored quantum circuit for Majorana fermions. In each time step, after unitary time evolution generated by $\hat{\mathcal{H}}_{\mathrm{Kitaev}}$ in Eq.~\eqref{eq:Hamiltonian}, we weakly measure the Majoranas in the brickwork manner with Kraus operators in Eq.~\eqref{eq:Kraus}.
(b) Bulk topological invariant $\chi$ for the monitored system, obtained from the fermion parities $P^{\mathrm{PBC}/\mathrm{APBC}}$ for circuits under PBC/APBC.
The APBC involves the boundary measurement operators with twisted outcomes with respect to the PBC.
(c) The measurement-induced phases characterized by the Lyapunov gap, the existence of the edge modes, and our topological invariant.
The topological (trivial) gapped phase exhibits the presence (absence) of edge modes and $\chi=-1$ ($\chi=+1$).
The critical gapless phase is characterized by the dynamical behavior of $\chi$.
}
\label{fig:Circuit}
\end{figure}
In this Letter, we discover gapless edge modes and a bulk topological invariant that characterize measurement-induced phase transitions of monitored Majorana circuits and discuss their bulk-edge correspondence. Introducing effective Hamiltonians obtained through the Lyapunov analysis, we reveal
the presence (absence) of edge-localized zero modes inside the bulk gap in the topological (trivial) area-law phase.
We also construct a bulk topological invariant based on the fermion parity, 
extending the method in equilibrium systems to the monitored setting; this is accomplished by introducing a unique boundary condition with twisted measurement outcomes.
The invariant sharply distinguishes these gapped area-law phases, featuring the bulk-edge correspondence.
Moreover, we show that the critical entanglement phase corresponds to a gapless phase concerning the Lyapunov spectrum.
While the bulk-edge correspondence is obscured in this phase, we demonstrate that the topological invariant can still \textit{dynamically} characterize the two gapped and gapless phases. 
While we mainly focus on free fermionic systems in this Letter, our methodologies can be extended to many-body interesting systems and thus offer a general framework for the study of measurement-induced topological phases.
Our result is summarized in Fig.~\ref{fig:Circuit}.

\paragraph*{Model and Lyapunov analysis--}

We consider a $(1+1)$-dimensional quantum circuit acting on $2L$ Majorana fermions [see Fig.~\ref{fig:Circuit}(a)], consisting of repeated applications of three elementary steps: (i) unitary time evolution $\hat{\mathcal{U}}$, (ii) measurements of all Majorana pairs on odd bonds, and (iii) measurements of all Majorana pairs on even bonds.
These operations update $\ket{\psi_t}$ to $\ket{\psi_{t+1}}$ in a single time step.
To be more precise, the unitary evolution is given by the Kitaev chain Hamiltonian, $\hat{\mathcal{U}} = e^{-i\hat{\mathcal{H}}_\mathrm{Kitaev}}$, with
\begin{align}
\label{eq:Hamiltonian}
\hat{\mathcal{H}}_\mathrm{Kitaev} = iJ \sum_{\ell=1}^{2L-1} \gamma_{\ell} \gamma_{\ell+1} +iJ' \gamma_{2L} \gamma_1, 
\end{align}
where $\gamma_\ell$ are Majorana fermions obeying $\{\gamma_\ell, \gamma_{\ell'} \}=2\delta_{\ell \ell'}$ and $J,J'$ are real constants.
The measurements of parities for neighboring Majorana pairs $i\gamma_{\ell}\gamma_{\ell+1}$ on odd or even bonds with an outcome $s=\pm 1$ are described by the Kraus operators~\cite{foot1}
,
\begin{align}
\label{eq:Kraus}
\hat{\mathcal{K}}^o_j(s) = \frac{e^{-is\theta_o \gamma_{2j-1} \gamma_{2j}}}{\sqrt{2\cosh (2\theta_o)}}, \quad
\hat{\mathcal{K}}^e_j(s) = \frac{e^{-is\theta_e \gamma_{2j} \gamma_{2j+1}}}{\sqrt{2\cosh (2\theta_e)}}.
\end{align}
which are weak measurements of the strength $\theta_{e/o} = \tanh^{-1}(\mu_{e/o})$.
The Kraus operator reduces to a projective measurement for $\theta_{o/e}\rightarrow\infty$ ($\mu_{o/e}\rightarrow1$) or to the identity operation for $\theta_{o/e}\rightarrow0$ ($\mu_{o/e}\rightarrow0$).
Experimental setups for implementing the Majorana parity measurements are discussed in \cite{Kells23, Lutchyn18}.
We set $\mu_e = 1-\mu_o$ hereafter to simply interpolate between the limits $(\mu_o, \mu_e)=(0,1)$ and $(1,0)$, where the two distinct area-law entangled phases are certainly realized.
Given a (un-normalized) state $\ket{\psi}$, each outcome $s$ is obtained with the Born probability,
$
p^{e/o}_j(s) = \bra{\psi}\hat{\mathcal{K}}_j^{e/o}(s)^\dagger \hat{\mathcal{K}}_j^{e/o}(s)\ket{\psi}/{\langle \psi | \psi \rangle}.
$
In the following, we use three boundary conditions: open (OBC), periodic (PBC), and antiperiodic (APBC). 
For the unitary evolution $\hat{\mathcal{U}}$, the boundary condition is simply specified by $J'$: $J'=0, J,$ and $-J$ for the OBC, PBC, and APBC, respectively.
For the measurements, the OBC means that we discard $\hat{\mathcal{K}}^e_L(s)$ from the circuit, while we need a careful definition of the APBC, as explained later.

At the time $t=T$, an initial state $\ket{\psi_0}$ is evolved by the Kraus operator labeled by {a} sequence of outcomes $\bm{s}=\{ \bm{s}_1, \ldots, \bm{s}_T \}$ with $\bm{s}_t = \{ s_{1,t}, \ldots, s_{2L,t} \}$, 
\begin{align}
    \hat{\mathcal{K}}_T(\bm{s}) = \prod_{t=1}^T \left(
    \left(\prod_{j}\hat{\mathcal{K}}_j^e(s_{2j,t})\right)
    \left(\prod_{j}\hat{\mathcal{K}}_j^o(s_{2j-1,t})\right)
    \hat{\mathcal{U}}
    \right).
    \label{eq:many-body Kraus operator}
\end{align}
Since both unitary evolution and measurements are bilinear in Majorana fermions, this circuit maps a fermionic Gaussian state to another Gaussian state \cite{Bravyi05, Turkeshi21, Surace22, Fidkowski21, Piccitto22, Fava23, Ravindranath23}.
A fermionic Gaussian state is completely characterized by the Majorana covariance matrix $\Gamma_t$ whose element reads $(\Gamma_t)_{\ell \ell'} = (i/2)\bra{\psi_t}[\gamma_\ell, \gamma_{\ell'}]\ket{\psi_t}/\langle\psi_t|\psi_t\rangle$.
Time evolution of the (un-normalized) Gaussian state is encoded in the transformation of the Majorana fermions, $\vec{\gamma} \to \hat{\mathcal{K}}_T^\dagger(\bm{s}) \vec{\gamma} [\hat{\mathcal{K}}_T^\dagger(\bm{s})]^{-1} = K_T(\bm{s}) \vec{\gamma}$, where
\begin{align}
    K_T(\bm{s}) = \prod_{t=1}^T \left(
    e^{-i\Theta^e(\bm{s}_t)}
    e^{-i\Theta^o(\bm{s}_t)}
    e^{H_\mathrm{Kitaev}}
    \right).
    \label{eq:single-particle Kraus operator}
\end{align}
Here, $H_\mathrm{Kitaev}$ and $\Theta^{e/o}(\bm{s}_t)$ are real antisymmetric matrices defined through $\hat{\mathcal{H}}_\mathrm{Kitaev} = i\vec{\gamma}^\mathsf{T} H_\mathrm{Kitaev}\vec{\gamma}/4$ and $\prod_j \hat{\mathcal{K}}^{e/o}_j(s_{2j/2j-1,t}) \propto \exp[{i\vec{\gamma}^\mathsf{T}\Theta^{e/o}(\bm{s}_t)}\vec{\gamma}/4]$ with $\vec{\gamma} = (\gamma_1, \ldots, \gamma_{2L})^\mathsf{T}$ (see Supplemental Material~\cite{supp}).

Since this circuit $K_T(\bm{s})$ is a product of random matrices, we can perform the Lyapunov analysis \cite{Zabalo22,Mochizuki23, Kumar24, Luca24, Mochizuki24, Bulchandani24, Xiao24a}.
Provided that Oseledec's theorem \cite{Crisanti93, Ershov98, Ginelli13} holds, the Lyapunov spectrum $z_\ell$ does not depend on the sequence of outcomes $\bm{s}$. 
It is also related to the energy spectrum of the effective Hamiltonian,
\begin{align}
    H_{\textrm{eff},T}(\bm{s}) = -\frac{i}{2T}\ln[K_T(\bm{s}) K_T^\dagger(\bm{s})],
\end{align}
and the corresponding many-body Hamiltonian $\hat{\mathcal{H}}_{\textrm{eff},T}(\bm{s}) = -i\Vec{\gamma}^\mathsf{T} H_{\textrm{eff},T}(\bm{s}) \Vec{\gamma}/4$. 
Since $H_{\textrm{eff},T}(\bm{s})$ is real antisymmetric, its eigenvalues $iz_{\ell,T}(\bm{s}) \in i\mathbb{R}$ come in pairs, $z_{2j-1,T}(\bm{s}) = -z_{2j,T}(\bm{s})$.
This defines the single-particle energy spectrum $z_{\ell,T}(\bm{s})$, which in the asymptotic limit coincides with the Lyapunov spectrum: $z_\ell = \lim_{T \to \infty} z_{\ell,T}(\bm{s})$. 
We below arrange them as $0 \leq z_1 = -z_2 \leq \cdots \leq z_{2L-1} = -z_{2L}$.
We can also compute the corresponding Lyapunov vectors $\vec{w}_{\ell,T}(\bm{s})$.
Written in the matrix form $\widetilde{O}_T(\bm{s}) = (\vec{w}_{1,T}(\bm{s}), \ldots, \vec{w}_{2L,T}(\bm{s}))$, they give in the asymptotic limit an orthogonal matrix that brings $H_{\textrm{eff},T}(\bm{s})$ into the standard form, 
\begin{align}
\lim_{T \to \infty} \widetilde{O}_T^\mathsf{T}(\bm{s}) H_{\textrm{eff},T}(\bm{s}) \widetilde{O}_T(\bm{s}) = \bigoplus_{j=1}^L \begin{pmatrix}{0}& z_{2j-1} \\ -z_{2j-1} &{0} \end{pmatrix}.
\label{eq:standard form of Heff}
\end{align}
In practice, we construct $\widetilde{O}_T(\bm{s})$ through computing Lyapunov vectors corresponding to the non-negative Lyapunov spectrum based on the complex fermion representation, which is equivalent to the Majorana representation~\cite{supp}.
In the following, the initial state of the trajectories is fixed to be a vacuum state $\ket{\psi_0} = \ket{0}$, whereas the Lyapunov analysis itself is performed with random initial vectors $\vec{w}_{\ell,0}$ for a given trajectory.
We only look at a single trajectory specified by a typical sequence of $\bm{s}$ and consider a temporal average of quantities in the long-time limit unless otherwise mentioned~\cite{supp}.
Note that, in contrast to the conventional setup of products of random matrices, the Born rule unique to monitored systems makes the existence or convergence of the Lyapunov spectrum a mathematically nontrivial problem, which has only recently been addressed in Refs.~\cite{Benoist19, Benoist21}.

\begin{figure*}
\includegraphics[width=1.0\textwidth]{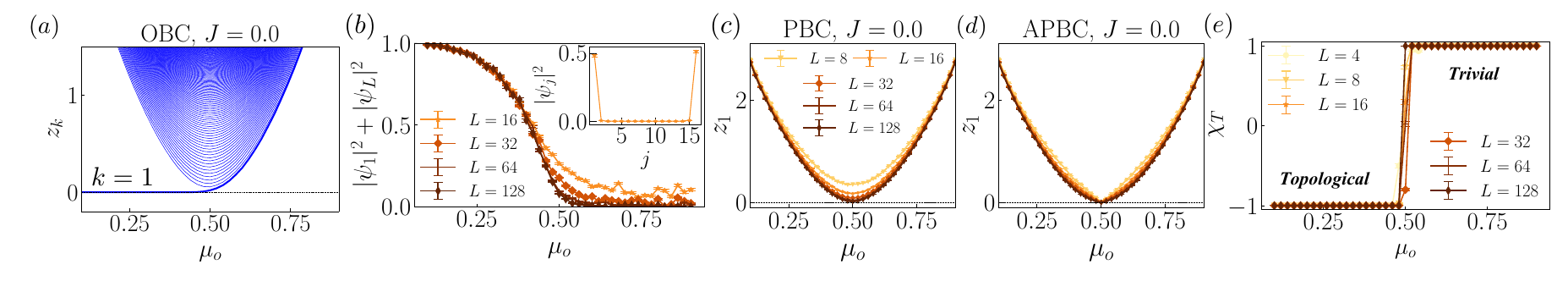}
\caption{
(a) Non-negative single-particle Lyapunov spectra for $L=128$ and (b) sum of the squared amplitudes of the spinor wave function corresponding to the lowest energy $z_1$ at the edges for the measurement-only circuit with OBC. 
The inset in (b) shows the spatial distribution of the squared amplitudes of the spinor wave function at $t=2L$ averaged over $1000$ different trajectories for $L=16$ and $\mu_o = 0.1$.
(c),(d) Lowest non-negative single-particle Lyapunov spectrum for the measurement-only circuit with (c) PBC and (d) APBC.
(e) Topological invariant after a sufficiently long time.
}
\label{fig:MeasOnlyLyapunov}
\end{figure*}
\paragraph*{Lyapunov edge modes in measurement-only circuits--}

We first perform the Lyapunov analysis for the measurement-only circuit ($J=J'=0$).
In this case, a direct transition at $\mu_o\simeq 0.5$ occurs from a topological to a trivial area-law phase, as numerically checked by the topological entanglement entropy and bipartite mutual information~\cite{supp}, following the approach in \cite{Kells23, Nehra24, Pan24}.

We find that the topological transition between the two area-law phases can be viewed as the transition between two bulk-gapped phases for the Lyapunov spectrum, and the bulk gap closes once at the transition.
Remarkably, we reveal that the topological (trivial) area-law phase is distinguished by the presence (absence) of Majorana zero modes within the bulk gap.
Figure~\ref{fig:MeasOnlyLyapunov}(a) shows the non-negative single-particle Lyapunov spectrum $z_k$ against $\mu_o$ for the measurement-only circuit under the OBC for $L=128$.
Importantly, $z_1$ takes a finite value for $\mu_o>0.5$, while it vanishes for $\mu_o\leq0.5$.
Vanishing of $z_1$ under the OBC in the topological area-law phase implies the many-body Lyapunov gap closing for $\hat{\mathcal{H}}_{\textrm{eff},T}(\bm{s})$, which is reminiscent of the appearance of Majorana zero modes for the static Kitaev chain \cite{Kitaev01}.

We next show that the above zero modes are spatially localized edge modes by analyzing the spinor wave function $\vec{\psi}_T(\bm{s})$ \cite{Mbeng24} with its element
$[\vec{\psi}_{T}(\bm{s})]_j = \sqrt{\sum_{a=1,2}\sum_{b=2j-1,2j}[\vec{w}_{a,T}(\bm{s})]_{b}^2 / 2}$ corresponding to the lowest energy $z_{1,T}(\bm{s}) \simeq z_1$.
In Fig.~\ref{fig:MeasOnlyLyapunov}(b), we plot the long-time average of the sum of the squared amplitudes for $\vec{\psi}_T(\bm{s})$ at both edges, $|[\vec{\psi}_T(\bm{s})]_1|^2+|[\vec{\psi}_T (\bm{s})]_L|^2$, against $\mu_o$.
Deep inside the topological phase ($\mu_o \ll 0.5$), it takes a value close to $1$ regardless of the system size, signaling localization of the zero modes near the edges.
In contrast, it decreases toward a value $2/L$ in the trivial phase, indicating that the Majorana modes delocalize over the whole lattice on average.
The localization of the zero modes in the topological phase is much more visible if we directly look at the spatial profile of the $|[\vec{\psi}_T(\bm{s})]_j|^2$. 
The inset of Fig.~\ref{fig:MeasOnlyLyapunov}(b) shows (trajectory-averaged) values of $|(\vec{\psi}_T(\bm{s}))_j|^2$ for $L=16$ and $\mu_o=0.1$. 
These results indicate that the Lyapunov spectrum clearly reveals edge-localized Majorana zero modes in the topological area-law phase.

\paragraph*{Topological invariant and bulk-edge correspondence--}
We now introduce a bulk topological invariant and relate it with the presence (absence) of the Lyapunov edge modes in the topological (trivial) area-law phase.
Since the circuit $\hat{\mathcal{K}}_T(\bm{s})$ conserves the fermion parity $\hat{P} = \prod_{j=1}^L i\gamma_{2j-1} \gamma_{2j}$, each eigenstate of the effective Hamiltonian $\hat{\mathcal{H}}_{\textrm{eff},T}(\bm{s})$ has a definite parity unless it is degenerate.
We then define the topological invariant as a difference between the fermion parities computed for the ground state under the PBC and that under the APBC.
While this is analogous to the procedure for static Kitaev chains \cite{Kitaev01, Beenakker15, Kawabata17}, special care is required to determine the APBC because the trajectories depend on measurement outcomes. 
Namely, the APBC is here defined \textit{with respect to the PBC}; we generate a circuit under the PBC with a sequence of measurement outcomes $\bm{s}$. 
Then the circuit under the APBC is defined by flipping only the outcomes $s_{2L,t}$ for the boundary Majorana pairs, $\hat{\mathcal{K}}^e_L(s_{2L,t}) \to \hat{\mathcal{K}}^e_L(-s_{2L,t})$, with $\hat{\mathcal{U}}$ under the APBC.
Using $\mathcal{\hat{H}}_{\mathrm{eff},T}^\mathrm{PBC/APBC}(\bm{s})$ defined through the above prescription, we introduce the topological invariant as the product of their ground states' parities.
Since the parity can be computed as the sign of the Pfaffian of the single-particle Hamiltonian for noninteracting systems, which is known as Kitaev's formula~\cite{Kitaev01}, the topological invariant for monitored Majorana circuits is given by ${Q_T(\bm{s}) = \mathrm{sgn}\left(\mathrm{Pf}[{H}_{\mathrm{eff},T}^\mathrm{PBC}(\bm{s})] \mathrm{Pf}[{H}_{\mathrm{eff},T}^\mathrm{APBC}(\bm{s})]\right)}$.
However, the explicit form of $H_{\textrm{eff},T}(\bm{s})$ is numerically intractable for large $T$.
Thus, we instead compute $\chi_T(\bm{s})$ defined by
\begin{align}
   \label{eq:TopoInvDet}
   \chi_T(\bm{s}) &= P_T^\textrm{PBC}(\bm{s}) P^\textrm{APBC}_T(\bm{s}), \\
   P^{\mathrm{PBC/APBC}}_T(\bm{s}) &= \det[\widetilde{O}^\textrm{PBC/APBC}_T(\bm{s})].
\end{align}
Since $\chi_T(\bm{s})$ approximates and converges to $Q_T(\bm{s})$ for $T \to \infty$, we will also call $\chi_T(\bm{s})$ as the topological invariant. 
We note that $\widetilde{O}^\textrm{PBC/APBC}_T(\bm{s})$ is not an orthogonal matrix for general $T$.
Note also that the Lyapunov spectrum and topological invariant have been discussed in Refs.~\cite{Venn23, Behrends24} for a deterministic nonunitary circuit, while these works did not consider the monitored system determined by the Born rule.

The experimental overhead to compute the topological invariant is exponential in the number of measurements, since the measurement outcomes in the circuit under the APBC are required to be postselected with respect to those in the circuit under the PBC.
In general, we have similar postselection problems to observe the measurement-induced phenomena \cite{Fisher23}.
While there are several works trying to overcome this difficulty \cite{Gullans20PRL, Noel22, Hoke23, Agrawal24, Kamakari24, Barratt22, Dehghani23, Garratt23, Garratt24, Iadecola22, Buchhold22, Yamamoto25} and brute-force postselection is possible for small-size systems~\cite{Koh23}, a definitive method to circumvent postselection problems remains unclear and is reserved for future work.

In Figs.~\ref{fig:MeasOnlyLyapunov}(c) and \ref{fig:MeasOnlyLyapunov}(d), we show the lowest single-particle Lyapunov spectrum $z_1$ for the measurement-only circuit under the PBC and APBC, respectively.
It takes a finite value for both cases, implying a finite bulk Lyapunov gap, within each area-law phase.
As shown in Fig.~\ref{fig:MeasOnlyLyapunov}(e), the topological invariant $\chi_T(\bm{s})$ clearly separates the topological and trivial area-law phases.
Note that these results are qualitatively independent of the measurement outcomes $\bm{s}$.
In fact, $P^{\mathrm{PBC}}_T(\bm{s})=1$ is satisfied in the whole parameter region, since the gap does not close at the transition $\mu_o \simeq 0.5$ for the PBC due to finite-size splitting $z_1 \sim 1/L$.
On the other hand, $P^{\mathrm{APBC}}_T(\bm{s})$ abruptly changes from $-1$ to $+1$ across the transition point for the APBC, implying an exact gap closing near $\mu_o=0.5$.
This results in the observed transition in $\chi_T(\bm{s})$.
The edge-localized zero modes under the OBC in Figs.~\ref{fig:MeasOnlyLyapunov}(a) and \ref{fig:MeasOnlyLyapunov}(b) and the topological invariant in Fig.~\ref{fig:MeasOnlyLyapunov}(e) indicate that the bulk-edge correspondence applies even to nonequilibrium phases under temporally random measurements, if the bulk gap is open. 

\paragraph*{Critical gapless phase with additional unitary--}

\begin{figure*}
\includegraphics[width=0.97\textwidth]{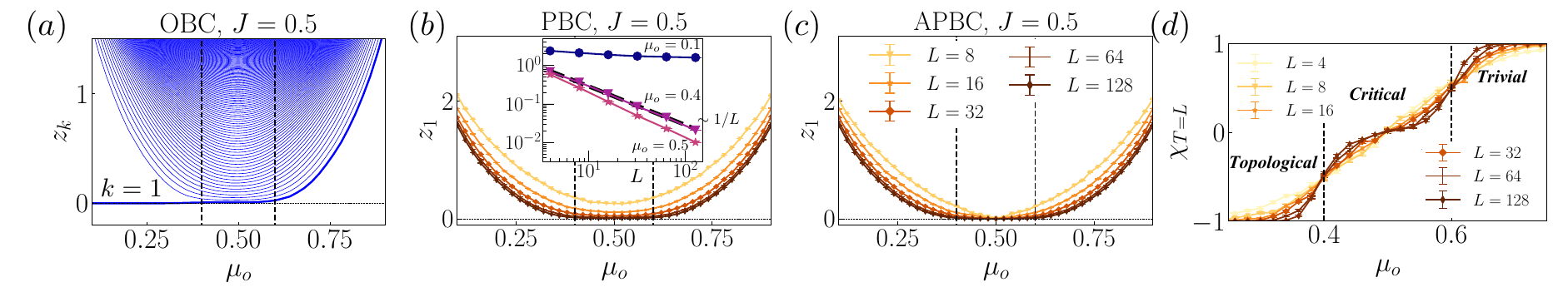}
\caption{
(a) Non-negative single-particle Lyapunov spectrum for the monitored circuit with $J=0.5$ under OBC for $L=128$.
(b),(c) Lowest non-negative spectrum $z_1$ under (b) PBC and (c) APBC.
The inset in (b) shows the system-size dependence of $z_1$ at $\mu_o=0.1, 0.4$, and $0.5$.
(d) Topological invariant at $T=L$ averaged over $1000$ different trajectories. 
The Lyapunov spectrum and topological invariant distinguish the topological area-law, critical sub-volume-law, and trivial area-law phases at $\mu_o \simeq 0.4$ and $0.6$, as indicated by the dashed vertical lines.
}
\label{fig:UnitaryLyapunov}
\end{figure*}

We now study the monitored circuit with unitary dynamics by the Kitaev Hamiltonian $\mathcal{\hat{H}}_\mathrm{Kitaev}$ and discuss how the Lyapunov spectrum and the topological invariant behave.
Reference~\cite{Fava23} considered a continuous-time version of our model to find three different entanglement phases as $\mu_o$ increases.
Specifically, the unitary part leads to a critical phase whose entanglement scales as $S\sim (\ln L)^2$ between the topological and trivial area-law phases (see also \cite{Nahum20, Bao21, Sang21PRXQ, Jian22, Jian23, Loio23, Klocke23, Kells23, Merritt23, Pan24}).
In our circuit model, we numerically find three entanglement phases separated by two transition points $\mu_o\simeq 0.4$ and $\simeq 0.6$ for $J=0.5$, using the topological entanglement entropy and bipartite mutual information~\cite{supp}.

To discuss the topology of the three phases, we first investigate the Lyapunov spectrum in the circuit with the OBC, finding that the critical phase corresponds to a gapless phase without Majorana zero modes.
Figure~\ref{fig:UnitaryLyapunov}(a) shows the $\mu_o$ dependence of the non-negative single-particle Lyapunov spectrum.
As observed in the measurement-only case, the lowest mode $z_1$ becomes gapless (gapped) in the topological (trivial) phase.
In the critical phase for $0.4 \lesssim \mu_o \lesssim 0.6$, not only the lowest mode $z_1$ but also higher modes $z_{k(\neq 1)}$ appear to become gapless, implying the closing of the bulk Lyapunov gap. 
Interestingly, the $L$ dependence of $z_1$ seems different between the topological area-law phase and the critical phase; $z_1$ decays faster than power law in the topological area-law phase ($\mu_o\lesssim0.4$), while it appears to decay in $1/L$ for our available system sizes in the critical phase ($0.4\lesssim\mu_o\lesssim 0.6$)~\cite{supp}.

The above observation of the bulk spectrum is also confirmed from the lowest spectrum $z_1$ under the PBC and APBC.
Indeed, as shown in Figs.~\ref{fig:UnitaryLyapunov}(b) and \ref{fig:UnitaryLyapunov}(c), $z_1$ goes to zero for $0.4\lesssim \mu_o\lesssim 0.6$, whereas the bulk gap is open for other $\mu_o$. 
While the scaling form of $z_1$ right at the transitions $\mu_o \simeq 0.4$ and $0.6$ is close to $1/L$, it appears to decay faster than $1/L$ inside the critical phase (except the APBC circuit at $\mu_o=0.5$, where exact gap closing is anticipated nearby)~\cite{supp}.

The above results for the spectral gaps indicate that the bulk-edge correspondence is obscured for the gapless phase if we consider $\chi_\infty\equiv\lim_{T\rightarrow\infty}\chi_T(\bm{s})$ for finite $L$, which is confirmed to be convergent irrespective of $\bm{s}$.
That is, since $z_1$ can only be exactly zero near $\mu_o=0.5$ for the APBC, the asymptotic topological invariant can switch the sign only at this point for finite system sizes.
Indeed, we have numerically observed that $\chi_T$ flows toward $-1$ for $\mu_o<0.5$ and $+1$ for $\mu_o>0.5$ as $T$ increases~\cite{supp}.
In contrast, as discussed above, the Majorana zero modes seem absent even for $0.4 \lesssim \mu_o < 0.5$.
Therefore, while the critical phase is unstable in static Majorana chains and is thus unique to monitored Majorana chains, the bulk-edge correspondence is obscured since it corresponds to the gapless phase. 
This is reminiscent of the fact that the topological phase is usually well defined in gapped phases in static systems.

Remarkably, however, we find that the topological invariant \textit{dynamically} characterizes the different phases, including the critical gapless phase. 
Specifically, we consider the sample-averaged invariant $\chi_T$ for the timescale $T=\mathcal{O}(L)$.
Figure~\ref{fig:UnitaryLyapunov}(d) shows $\chi_{T=L}$ for different system sizes, from which we find three crossings with $\mu_o\simeq 0.4, 0.5,$ and $0.6$.
The locations of the first and last crossings coincide well with the transitions between the area-law gapped phases and critical gapless phase.
Notably, this characterization is attributed to the divergent relaxation time of the invariant in the gapless phase.
If we denote the timescale as $\tau_\mathrm{relax}$, it is at least longer than $\sim 1/z_1$, which increases faster than $L$ inside the gapless phase.
Therefore, $\chi_{T=L}$ will not converge and almost stay zero for the gapless phase.
This is contrasted to the gapped phases with $z_1=\mathcal{O}(L^0)$, where $\chi_{T}$ rapidly converges to $\pm 1$ for $T=\mathcal{O}(L)$.
Note that the crossing point at $\mu_o\simeq 0.5$ is different from the boundaries of the three phases. The above argument indicates that this crossing is a finite-size artifact, and $\chi_{T=L}$ will flatly become zero in the entire critical phase for $L\rightarrow\infty$.

\paragraph*{Summary and outlook--}
We have shown that monitored Majorana circuits in the gapped phases exhibit the bulk-edge correspondence by investigating the Lyapunov spectrum, gapless edge modes, and topological invariants based on the fermion parity. 
To define the topological invariant, we have used a unique methodology where the trajectory under the APBC is defined with respect to that under the PBC, according to twisted measurement outcomes at the boundary. 
We have also shown that the critical phase corresponds to the gapless phase in terms of the Lyapunov spectrum and is dynamically characterized by the topological invariant in a timescale $t=\mathcal{O}(L)$. 

Our study opens the way to explore topological features in monitored systems through the Lyapunov analysis.
An important future direction is to examine many-body systems, where our topological invariant based on the twisted measurement outcomes readily applies.
For example, it is intriguing to analyze monitored circuits with $Z_2 \times Z_2$ symmetry, for which measurements can stabilize an area-law phase with a symmetry-protected topological order of the cluster state \cite{Lavasani21, Bao21, Morral-Yepes23}. 
Such a topological area-law phase will be characterized by a (nearly) fourfold degenerate Lyapunov spectrum under the OBC from gapless edge modes and a topological invariant measuring the change of one $Z_2$ charge by twisting boundary measurement outcomes for another $Z_2$ symmetry.
Another promising direction is to extend our topological invariant to measurement-induced topological phases protected by $U(1)$ symmetry.

\textit{Note added}.---
While this work was in its final stage, we learned about a recent paper that investigates a similar issue \cite{Xiao24b}.

\begin{acknowledgments}
\textit{Acknowledgments}---
We thank Xhek Turkeshi, Henning Schomerus, Alessandro Romito, Amos Chan, Keiji Saito, Ken Shiozaki, and Takahiro Morimoto for valuable discussions and comments.
We thank Zhenyu Xiao and Kohei Kawabata for coordinating the submission of this work and Ref.~\cite{Xiao24b}.
H.O. is supported by RIKEN Junior Research Associate Program.
He also appreciates the hospitality of Lancaster University where a part of this investigation was done.
K. M. and R.H. are supported by JST ERATO Grant Number JPMJER2302, Japan. 
K.M. is supported by JSPS KAKENHI Grant No. JP23K13037.
R.H. is supported by JSPS KAKENHI Grant No. JP24K16982.
Y.F. is supported by JSPS KAKENHI Grants No. JP20K14402 and No. JP24K06897. 
Part of the computation has been performed using the facilities of the Supercomputer Center, the Institute for Solid State Physics, the University of Tokyo.
\end{acknowledgments}

\textit{Data availability}.---
Codes for the simulations and numerical data are shared in Ref.~\cite{zenodo}

\bibliography{MajoTopoMIPT}

\begin{thebibliography}{95}
\bibitem [{foo()}]{foot1}%
  \BibitemOpen
  \href@noop {} {\bibinfo  {journal} {Under the Jordan-Wigner transformation, the measurements of the parities for neighboring Majorana fermions on an odd bond $i\gamma_{2j-1}\gamma_{2j}$ and an even bond $i\gamma_{2j}\gamma_{2j+1}$ are mapped to measurements of $Z_j$ and $X_jX_{j+1}$, respectively.}\ }\BibitemShut {NoStop}%
\end{thebibliography}

\clearpage

\onecolumngrid
\setcounter{equation}{0}
\setcounter{figure}{0}
\renewcommand{\theequation}{S\arabic{equation}}
\renewcommand\thefigure{S\arabic{figure}} 

\begin{center}
{\large \textbf{Supplemental Material: Topology and Spectrum in Measurement-Induced Phase Transitions}}
\end{center}

\pagenumbering{arabic}

\section{Numerical details}
In this section, we provide a brief overview of the numerical simulation of our monitored Majorana circuits and outline the procedure for performing Lyapunov analysis.
In this Supplemental Material, we assume that the states are normalized at every time.

\subsection{Time evolution of the correlation matrix}
\label{subseq:SM:Time evolution of correlation matrix} 
While we focus on the covariance matrix of the Majorana fermions in the main text, we here consider the correlation matrix of complex fermions for stable numerical simulations. 
The correspondence between these expressions, along with the numerical methods, is elaborated in the following. 
As considered in the main text, we address a system with $2L$ Majorana modes represented by $\vec{\gamma}=(\gamma_1,\ldots,\gamma_{2L})^\mathsf{T}$, where $\{\gamma_\ell,\gamma_{\ell'}\}=2\delta_{\ell\ell'}$ is satisfied.
Alternatively, we can consider the annihilation and creation operators of complex fermions as $\vec{\phi}_c=(c_1,\ldots,c_L,c_1^\dagger,\ldots,c_L^\dagger)^\mathsf{T}$, where $c_j=(\gamma_{2j-1}+i\gamma_{2j})/2$ and $c_j^\dagger=(\gamma_{2j-1}-i\gamma_{2j})/2$.
These operators satisfy the canonical anticommutation relations $\{c_i,c_j\}=0$ and $\{c_i, c_j^\dagger\}=\delta_{ij}$.
The relationship between $\vec{\phi}_c$ and $\vec{\gamma}$ can be expressed as $\Omega\vec{\phi}_c=\vec{\gamma}$, where $\Omega$ is given by $\Omega=S\tilde{\Omega}$ with $2L\times2L$ matrices $S$ and $\tilde{\Omega}$ defined as 
\begin{align}
    S_{ij} = 
        \begin{cases}
        & \delta_{(i+1)/2,j} \:\:\textrm{if }i:\textrm{odd}\\
        & \delta_{L+i/2,j} \:\:\:\:\textrm{if }i:\textrm{even} \\
        \end{cases}
        ,
    \quad
    \textrm{and},
    \quad
    \tilde{\Omega}=\mqty(
    I_L & I_L \\
    -iI_L & iI_L
    ).
\end{align}
It is straightforward to verify that $\Omega\Omega^\dagger=\Omega^\dagger\Omega=2$ and unitarity of $\Omega/\sqrt{2}$.

All the correlation functions of any fermionic Gaussian state are fully characterized by the two-point covariance matrix $\Gamma$ whose elements are ${\Gamma_{ij}}=\frac{i}{2}\left\langle[\gamma_i,\gamma_j]\right\rangle$. 
Alternatively, the state can be described using the correlation matrix of complex fermions
\begin{align}
    C = \langle\vec{\phi}_c\vec{\phi}_c^\dagger\rangle
    =\mqty(G & F \\ -F^* & 1-G^\mathsf{T}),
\end{align}
where $G_{ij}=\langle c_ic_j^\dagger\rangle$ and $F_{ij}=\langle c_ic_j\rangle$.
The correlation matrix $C$ and the covariance matrix $\Gamma$ are related via $C=\Omega^\dagger(-i\Gamma+I)\Omega/4$.
We can interchangeably use these two expressions.

\subsubsection{Unitary dynamics}
The Kitaev chain Hamiltonian considered in the main text is given by
\begin{align}
    \hat{\mathcal{H}}_{\textrm{kitaev}} 
    &= iJ\sum_{\ell=1}^{2L-1}\gamma_l\gamma_{l+1} + iJ'\gamma_{2L}\gamma_1
    \label{eq:SM:Kitaev Majorana}
    \\
    &= -J\sum_{j=1}^{L-1}(c_j^\dagger c_{j+1} + c_j^\dagger c_{j+1}^\dagger + h.c.) - J'(c_L^\dagger c_1 + c_L^\dagger c_1^\dagger + h.c.) + J\sum_{j=1}^L(2c_j^\dagger c_j-1).
    \label{eq:SM:Kitaev complex fermion}
\end{align}
The single-particle Hamiltonians for complex and Majorana fermions, denoted as $\widetilde{H}_{\mathrm{Kitaev}}$ and $H_{\mathrm{Kitaev}}$, respectively, satisfy
\begin{align}
    \hat{\mathcal{H}}_{\mathrm{Kitaev}} = \vec{\phi}_c^\dagger \widetilde{H}_{\mathrm{Kitaev}}\vec{\phi}_c
    = \frac{i}{4}\vec{\gamma}^\mathsf{T}H_\mathrm{Kitaev}\vec{\gamma},
    \quad
    \textrm{with}
    \quad
    H_{\mathrm{Kitaev}}=-i\Omega \widetilde{H}_{\mathrm{Kitaev}}\Omega^\dagger
    \label{eq:Kitaev Hamiltonian}
\end{align}
Here, $\widetilde{H}_\mathrm{Kitaev}$ is a $2L\times2L$ Hermitian matrix, while $H_\mathrm{Kitaev}$ is a $2L\times2L$ real antisymmetric matrix, with elements derived from Eqs.~(\ref{eq:SM:Kitaev complex fermion}) and (\ref{eq:SM:Kitaev Majorana}), respectively.
The Heisenberg evolution of $\vec{\phi}_c$ by $\hat{\mathcal{U}}=e^{-i\hat{\mathcal{H}}_\mathrm{Kitaev}}$ becomes
\begin{align}
    \hat{\mathcal{U}}^\dagger{\vec{\phi}_c}\hat{\mathcal{U}}
    = U\vec{\phi}_c
    \quad
    \textrm{with}
    \quad
    U=e^{-2i{\widetilde{H}_{\mathrm{Kitaev}}}}.
\end{align}

For any Gaussian pure state, there exist $L$ operators that annihilate it, which means that the set of the annihilation operators characterizes the state~\cite{Fava23,Ravindranath23}.
Let $\bm{c}=(c_1,\ldots,c_L)^\mathsf{T}$ and $\bm{d}=(d_1,\ldots,d_L)^\mathsf{T}$ be the annihilation operators of the initial state $\ket{\psi_0}$ and a general state $\ket{\psi}$, respectively.
There exists a unitary operator $\hat{\mathcal{V}}$  that maps the initial state $\ket{\psi_0}$ to $\ket{\psi}$ as $\ket{\psi}=\hat{\mathcal{V}}\ket{\psi_0}$, provided that both states are Gaussian pure states.
Thus, $\bm{d}$ can be written as
\begin{align}
    \bm{d} = \hat{\mathcal{V}}\bm{c}\hat{\mathcal{V}}^\dagger = \mathbb{U}^\dagger\vec{\phi}_c,
\end{align}
with $2L\times L$ matrix $\mathbb{U}$, which is verified to be an isometry satisfying $\mathbb{U}^\dagger\mathbb{U}=I_L$ due to the canonical anticommutation relations of $\bm{d}$.
The matrix $\mathbb{U}\mathbb{U}^\dagger$ is a rank $L$ projector, while the correlation matrix $C$ is also a rank $L$ projector~\cite{Surace22}.
Since they satisfy
\begin{align}
    \mathbb{U}\mathbb{U}^\dagger
    = \mathbb{U}\bra{\psi}\bm{d}\bm{d}^\dagger\ket{\psi}\mathbb{U}^\dagger
    =\mathbb{U}\mathbb{U}^\dagger\bra{\psi}\vec{\phi}_c\vec{\phi}_c^\dagger\ket{\psi}\mathbb{U}\mathbb{U}^\dagger
    = (\mathbb{U}\mathbb{U}^\dagger)C(\mathbb{U}\mathbb{U}^\dagger),
\end{align}
they should be identical,
\begin{align}
    C = \mathbb{U}\mathbb{U}^\dagger.
\end{align}
The above formula implies that the general Gaussian pure state $\ket{\psi}$ is characterized only by the isometry $\mathbb{U}$.

After the Gaussian unitary evolution $\hat{\mathcal{U}}$, the evolved state $\ket{\psi'}=\hat{\mathcal{U}}\ket{\psi}$ is annihilated by $\bm{f}=(f_1,\ldots,f_L)^\mathsf{T}$ with
\begin{align}
    \bm{f} = \hat{\mathcal{U}}{\bm{d}}\hat{\mathcal{U}}^\dagger
    = \mathbb{U}^\dagger \hat{\mathcal{U}}\vec{\phi}_{c}\hat{\mathcal{U}}^\dagger
    = \mathbb{U}^\dagger U^\dagger\vec{\phi}_{c}
    = (U\mathbb{U})^\dagger\vec{\phi}_c.
\end{align}
The above equation implies that the time evolution is dictated by the change in the isometry $\mathbb{U}$, i.e., $\mathbb{U}'=U\mathbb{U}$ and $C'=\mathbb{U}'(\mathbb{U}')^\dagger$.

Setting the initial state as the vacuum state of $\bm{c}$, it is straightforward to verify that the isometry $\mathbb{U}_0$ corresponding to the initial state is given by
\begin{align}
    \mathbb{U}_0 = \mqty(I_L \\ 0_L).
\end{align}

\subsubsection{Non-unitary evolution caused by measurements}
\label{subsubsec:SM:Non-unitary evolution caused by measurements}
We now consider the non-unitary dynamics caused by measurements.
The measurements considered in the main text are given by the Kraus operators $\hat{\mathcal{K}}_j^{e/o}(s) \propto e^{\hat{\Theta}^{e/o}_j(s)}$ up to the normalization factor. 
Here, the Hermitian operators $\hat{\Theta}^{e/o}_j(s)$ explicitly take the forms, 
\begin{align}
    \hat{\Theta}^{o}_j(s) = -is\theta_o\gamma_{2j-1}\gamma_{2j} = -s\theta_o(2c_j^\dagger c_j - 1) 
    = \mqty(c_j^\dagger & c_j)\mqty(
    -s{\theta_o} & {0} \\
    {0} & s{\theta_o}
    )\mqty(c_j \\ c_j^\dagger)
\end{align}
for the measurements of the Majorana pairs on odd bonds, whereas
\begin{align}
    \hat{\Theta}^{e}_j(s) &= -is\theta_e\gamma_{2j}\gamma_{2j+1}
    = s\theta_e(c_j^\dagger c_{j+1} + c_j^\dagger c_{j+1}^\dagger + h.c.) \\
    &= \mqty(c_j^\dagger & c_{j+1}^\dagger & c_j & c_{j+1})
    \mqty(
    {0} & s{\theta_e}/2 & {0} & s{\theta_e}/2 \\
    s{\theta_e}/2 & {0} & -s{\theta_e}/2 & {0} \\
    {0} & -s{\theta_e}/2 & {0} & -s{\theta_e}/2 \\
    s{\theta_e}/2 &{0} & -s{\theta_e}/2 & {0}
    )
    \mqty(c_j \\ c_{j+1} \\ c_j^\dagger \\ c_{j+1}^\dagger)
\end{align}
for the measurements of the Majorana pairs on even bonds. 
Since these operators are quadratic in fermions, we can introduce a $2L \times 2L$ real symmetric matrix $\widetilde{\Theta}^{e/o}_j(s)$ and a $2L \times 2L$ real antisymmetric matrix $\Theta^{e/o}_j(s)$ through
\begin{align}
    \hat{\Theta}^{e/o}_j(s) = \vec{\phi}_c^\dagger\widetilde{\Theta}^{e/o}_j(s)\vec{\phi}_c = \frac{i}{4}\vec{\gamma}^\mathsf{T}\Theta^{e/o}_j(s)\vec{\gamma}.
    \label{eq:Theta eo}
\end{align}
The Heisenberg evolution of the complex fermions under measurement is given by
\begin{align}
    \hat{\mathcal{K}}_j^{e/o}(s)^\dagger\vec{\phi}_c(\hat{\mathcal{K}}_j^{e/o}(s)^\dagger)^{-1}
    = \hat{\mathcal{K}}_j^{e/o}(s)\vec{\phi}_c\hat{\mathcal{K}}_j^{e/o}(s)^{-1}
    = K_j^{e/o}(s)\vec{\phi}_c
    \quad
    \textrm{with}
    \quad K_j^{e/o}(s) = e^{-2{\widetilde{\Theta}_j^{e/o}(s)}}
\end{align}

We generically consider Gaussian non-unitary evolution $\ket{\psi'}=\hat{\mathcal{K}}\ket{\psi}/\|\hat{\mathcal{K}}\ket{\psi}\|$ caused by the measurement described above.
Let $\bm{c}$ and $\bm{d}$ denote the operators that annihilate the initial vacuum state $\ket{\psi_0}$ and a general state $\ket{\psi}$, respectively.
The operators annihilating the evolved state $\ket{\psi'}$ are given by $\tilde{\bm{f}}=\hat{\mathcal{K}}\bm{d}\hat{\mathcal{K}}^{-1}$.
However, while the operators $\tilde{f}_i$ satisfy $\{\tilde{f}_i,\tilde{f}_j\}=0$, it is found that $\{\tilde{f}_i, \tilde{f}_j^\dagger\}\neq\delta_{ij}$, which means that they are not canonical fermionic operators.

To address this issue, we aim to construct a unitary transformation that maps the pre-measurement state $\ket{\psi}$ to the post-measurement state $\ket{\psi'}$.
Expressing $\bm{d}$ as $\bm{d}=\mathbb{U}^\dagger\vec{\phi}_c$ and writing $\hat{\mathcal{K}}\vec{\phi}_c\hat{\mathcal{K}}^{-1}=K\vec{\phi}_c$, the operator $\tilde{\bm{f}}$ can be written as
\begin{align}
    \bm{\tilde{f}} = \hat{\mathcal{K}} \bm{d}\hat{\mathcal{K}}^{-1}  =\mathbb{U}^\dagger\hat{\mathcal{K}}\vec{\phi}_c\hat{\mathcal{K}}^{-1}
    =\mathbb{U}^\dagger K\vec{\phi}_c
    =(K\mathbb{U})^\dagger\vec{\phi}_c
    =(\tilde{\mathbb{U}}')^\dagger\vec{\phi}_c
\end{align}
where $\tilde{\mathbb{U}}'\equiv K\mathbb{U}$ with the Hermitian matrix $K$.
We now apply the thin QR decomposition to $\tilde{\mathbb{U}}'$, yielding $\tilde{\mathbb{U}}'=\mathbb{Q}R$ where $\mathbb{Q}$ is a $2L\times L$ isometry satisfying $\mathbb{Q}^\dagger\mathbb{Q}=I_L$, and $R$ is an $L\times L$ upper triangular matrix.
We then define new operators $\bm{f}$ as
\begin{align}
    \bm{f} = (R^\dagger)^{-1}\tilde{\bm{f}} = \mathbb{Q}^\dagger\vec{\phi}_c.
\end{align}
Since $f_j$ is a linear combination of $\tilde{f}_j$, the new operators $f_j$ also annihilate the state {$\ket{\psi'}$}.
Furthermore, as $\mathbb{Q}$ is an isometry, the canonical anticommutation relation $\{f_i, f_j^\dagger\}=\delta_{ij}$ holds.
This implies the existence of a Gaussian unitary operator $\hat{\mathcal{Q}}$ such that ${\ket{\psi'}}=\hat{\mathcal{Q}}\ket{\psi}$, with
\begin{align}
    \bm{f} = \hat{\mathcal{Q}}\bm{d}\hat{\mathcal{Q}}^\dagger = \mathbb{Q}^\dagger\vec{\phi}_c.
\end{align}
Thus, the correlation matrix can be updated as $C'=\mathbb{Q}\mathbb{Q}^\dagger$.

\subsubsection{Born probability}
We next provide a way to compute the Born probabilities based on the correlation matrix when the weak measurements are applied.
The Kraus operators corresponding to the weak measurements of the Majorana pairs on $j$th odd and even bonds are written by
\begin{align}
    \hat{\mathcal{K}}_j^{o}(s) = \frac{1}{\sqrt{2\cosh(2\theta_o)}}e^{-is\theta_o\gamma_{2j-1}\gamma_{2j}}
    \quad
    \textrm{and}
    \quad
    \hat{\mathcal{K}}_j^{e}(s) = \frac{1}{\sqrt{2\cosh(2\theta_e)}}e^{-is\theta_e\gamma_{2j}\gamma_{2j+1}},
\end{align}
respectively.
For the measurements on the odd bond, by using $e^{ix\gamma_{\mu}\gamma_{\nu}}=\cosh{x}+i(\sinh{x})\gamma_{\mu}\gamma_{\nu}$ and transforming Majorana fermions to complex fermions, we find that the Born probability of finding the outcome $s$ is given by
\begin{align}
    p_{j}^o(s) = \left\langle \left(\hat{\mathcal{K}}_j^{o}(s)\right)^\dagger\hat{\mathcal{K}}_j^{o}(s)\right\rangle
    = \frac{1}{2(1+\mu_o^2)}((s-\mu_o)^2 + 4s\mu_o\langle c_j c_j^\dagger\rangle)
    = \frac{1}{2(1+\mu_o^2)}((s-\mu_o)^2 + 4s\mu_o G_{jj}).
\end{align}
In the same way, for the measurements on the even bond, the Born probability of finding the outcome $s$ is given by
\begin{align}
    p_{j}^e(s) &= \left\langle \left(\hat{\mathcal{K}}_j^e(s)\right)^\dagger\hat{\mathcal{K}}_j^e(s)\right\rangle \nonumber \\
    &= \frac{1}{2(1+\mu_e^2)}(1+\mu_e^2+2s\mu_e(\langle c_j^\dagger c_{j+1}\rangle + \langle c_j^\dagger c_{j+1}^\dagger\rangle + \langle c_{j+1}^\dagger c_j\rangle + \langle c_{j+1}c_{j}\rangle) \nonumber \\
    &= \frac{1}{2(1+\mu_e^2)}(1 + \mu_e^2 - 4s\mu_e \mathrm{Re}[G_{j,j+1} + F_{j,j+1}]).
\end{align}

\subsection{Lyapunov spectrum and Oseledec's theorem}
Oseledec's multiplicative ergodic theorem~\cite{Crisanti93} guarantees that, for almost all stationary sequences $\bm{\omega}={\{ \omega_1, \ldots, \omega_T \}}$, random matrices $A_t(\omega_t)$ with $\overline{\ln\|A_t(\omega_t)\|}<\infty$, the Oseledec matrix
\begin{align}
    \Xi(\bm{\omega}) = \underset{T\rightarrow\infty}{\lim}\left[
    K_T(\bm{\omega}) K_T^\dagger(\bm{\omega})
    \right]^{-\frac{1}{2T}},
    \label{eq:SM:Oseledec matrix}
\end{align}
exists.
Here, the overline refers to the average over the probability distribution of $\bm{\omega}$, and $K_T(\bm{\omega}) = A_T(\omega_T) \cdots A_1(\omega_1)$ is a product of the matrices in the sequence $\bm{\omega}$.
A $D$-dimensional Oseledec matrix has $D$ positive eigenvalues $\{e^{z_k(\bm{\omega})}\}$.
The exponents $z_k(\bm{\omega})$ are called the Lyapunov spectrum, and the eigenvectors of $\Xi(\bm{\omega})$ are called the Lyapunov vectors.
If $\bm{\omega}$ is an ergodic sequence, the Lyapunov spectrum does not depend on $\bm{\omega}$, i.e., $z_k\equiv z_k(\bm{\omega})$, although the  Lyapunov vectors still depend on $\bm{\omega}$ in general.

As defined in Eq. (4) of the main text, the matrix product $K_T(\bm{s})$ corresponding to our circuit is given in the Majorana fermion basis by
\begin{align}
    K_T(\bm{s}) = \prod_{t=1}^T \left(
    e^{-i\Theta^{e}(\bm{s}_{t})}
    e^{-i\Theta^{o}(\bm{s}_{t})}
    e^{H_{\mathrm{Kitaev}}}
    \right),
\end{align}
where $e^{-i\Theta^{e}(\bm{s}_{t})} = \prod_j e^{-i\Theta_j^{e}(s_{2j,t})}$, $e^{-i\Theta^{o}(\bm{s}_{t})} = \prod_j e^{-i\Theta_j^{o}(s_{2j-1,t})}$ , and $H_{\mathrm{Kitaev}}$ and $\Theta_j^{e/o}(s)$ are defined through Eqs.~(\ref{eq:Kitaev Hamiltonian}) and (\ref{eq:Theta eo}), respectively.

In practice, for the numerical simulation, it is convenient to use the matrix product $\widetilde{K}_T(\bm{s})$ represented in the complex fermion basis,
\begin{align}
    \widetilde{K}_T(\bm{s}) = \prod_{t=1}^T \widetilde{M}_t(\bm{s}_t),
    \quad
    \widetilde{M}_t(\bm{s}_t) = \left(
        \left(\prod_j e^{-2\widetilde{\Theta}_j^{e}(s_{2j,t})}\right)
        \left(\prod_j e^{-2\widetilde{\Theta}_j^{o}(s_{2j-1,t})}\right)
        e^{-2i\widetilde{H}_{\mathrm{Kitaev}}}
    \right).
\end{align}
We find that the condition $\overline{\ln\|A_t(\omega_t)\|}<\infty$ for Oseledec's theorem is satisfied as long as the measurements are not projective.
The logarithm of the operator norm of the matrix describing the unitary evolution is given by $0$.
Additionally, the logarithm of the operator norm of the matrix describing each variable-strength measurement is given by $2\theta_{e/o}-(\ln(2\cosh 2\theta_{e/o}))/2$, which is finite as long as $\mu_{e/o} =\tanh(\theta_{e/o})< 1$.

It is important to note that the single-particle Lyapunov spectrum for the monitored Majorana circuit comes in plus/minus pairs $0\leq z_{2j-1,T}(\bm{s})=-z_{2j,T}(\bm{s})$ due to the particle-hole symmetry of the effective Hamiltonian;
it is given by  $\widetilde{H}_{\mathrm{eff},T}(\bm{s}) = \ln\widetilde{\Xi}_T(\bm{s})$ with $\widetilde{\Xi}_T(\bm{s})=[\widetilde{K}_T(\bm{s})(\widetilde{K}_T(\bm{s}))^\dagger]^{-1/2T}$ in the complex fermion basis, which is connected to the effective Hamiltonian $H_{\mathrm{eff},T}(\bm{s})$ in the Majorana basis as $H_{\mathrm{eff},T}(\bm{s})=-i\Omega\widetilde{H}_{\mathrm{eff,T}}(\bm{s})\Omega^\dagger$.
The effective Hamiltonian $\widetilde{H}_{\mathrm{eff},T}(\bm{s})$ is diagonalized by a unitary matrix $W_T(\mathbf{s})$,
\begin{align}
    W_T^\dagger(\bm{s}) \widetilde{H}_{\mathrm{eff},T}(\bm{s}) W_T(\bm{s})
    = \mathrm{diag}(z_{1,T}(\bm{s}), z_{3,T}(\bm{s}), \ldots,z_{L-1,T}(\bm{s}),z_{2,T}(\bm{s}), z_{4,T}(\bm{s}), \ldots,z_{2L,T}(\bm{s})),
    \label{eq:SM:Def of Lyapunov vectors}
\end{align}
where $0\leq z_{1,T}(\bm{s})=-z_{2,T}(\bm{s})\leq\cdots \leq z_{2L-1,T}(\bm{s})=-z_{2L,T}(\bm{s})$.
Here, $z_{k,T}(\bm{s})$ computed in the complex fermion basis are the same as the $z_{k,T}(\bm{s})$ in the Majorana fermion basis in the main text.
Hence, we compute only the non-negative half of the Lyapunov spectrum along with the corresponding Lyapunov vectors, while the remaining half is determined by leveraging the particle-hole symmetry.

In general, however, it is numerically hard to compute the Oseledec matrix by directly multiplying random matrices due to numerical overflow~\cite{Ginelli13}.
To overcome this difficulty, we use a technique based on QR decomposition in our numerical calculation~\cite{Crisanti93, Ershov98, Ginelli13}.
We first prepare a $2L\times L$ random matrix $\mathbb{W}'_0$ whose elements are chosen from the complex Gaussian distribution and apply a thin QR decomposition to it as $\mathbb{W}'_0=\mathbb{Q}_0R_0$ and set $\mathbb{W}_0=\mathbb{Q}_0$.
Then, we repeat the following procedure:
\begin{enumerate}
    \item Apply the circuit operators $\widetilde{M}_t(\bm{s}_t)$ in one time step to $\mathbb{W}_{t-1}(\bm{s})$ as $\mathbb{W}'_t(\bm{s}) = \widetilde{M}_t(\bm{s}_t)\mathbb{W}_{t-1}(\bm{s})$.
    \item Apply QR decomposition to $\mathbb{W}'_t(\bm{s})$ as $\mathbb{W}'_t(\bm{s})=\mathbb{Q}_t(\bm{s})R_t(\bm{s})$, and set $\mathbb{W}_t(\bm{s})=\mathbb{Q}_t(\bm{s})$.
    \item Store the diagonal elements $\{(R_t(\bm{s}))_{jj}\}$ of $R_t(\bm{s})$.
\end{enumerate}
We note that $\mathbb{Q}_t(\bm{s})$ used here are not related to $\mathbb{Q}$ used in the Sec.~\ref{subsubsec:SM:Non-unitary evolution caused by measurements}.
Then, the snapshot Lyapunov spectrum at time $t=T$ is given by
\begin{align}
    \widetilde{z}_{2j-1,T}(\bm{s}) = \frac{1}{T}\sum_{t=1}^T\ln (R_t(\bm{s}))_{jj},
    \label{eq:SM:snapshot Lyapunov spectrum}
\end{align}
which converges to the non-negative Lyapunov spectrum $z_{2j-1,T}(\bm{s})$ as $T\to\infty$.
The corresponding snapshot Lyapunov vectors at $t=T$ are constructed by
\begin{align}
    \widetilde{W}_T(\bm{s}) = \mqty(
    \mathbb{W}_{T}^u(\bm{s}) & (\mathbb{W}_{T}^d(\bm{s}))^* \\ 
    \mathbb{W}_{T}^d(\bm{s}) & (\mathbb{W}_{T}^u(\bm{s}))^*
    ),
\end{align}
where $\mathbb{W}_t^u(\bm{s})$  ($\mathbb{W}_t^d(\bm{s})$) is the submatrix of $\mathbb{W}_t(\bm{s})$ consisting of rows ranging from 1 to $L$ (from $L+1$ to  $2L$) and all columns. 
This is because the effective Hamiltonian for complex fermions satisfies the particle-hole symmetry,
\begin{align}
    \Sigma_x\widetilde{H}_{\mathrm{eff},T}^*(\bm{s})\Sigma_x^{-1}=-\widetilde{H}_{\mathrm{eff},T}(\bm{s}),\ \ 
    \Sigma_x=\left(\begin{array}{cc}
     0_L  & I_L \\
      I_L  & 0_L
    \end{array}\right),
\end{align}
and thus the Lyapunov vectors corresponding to negative Lyapunov exponents can be constructed from those with positive exponents.
The matrix $\widetilde{O}_T(\bm{s})$ in the main text is obtained by $\widetilde{O}_T(\bm{s})=\Omega\widetilde{W}_T(\bm{s})\Omega^\dagger/4$.
Note that, while $\widetilde{W}_T(\bm{s})$ is not unitary for general $T$, it asymptotically becomes the unitary matrix $W_T(\bm{s})$ of the Lyapunov vectors that diagonalizes the effective Hamiltonian $\widetilde{H}_{\mathrm{eff},T}(\bm{s})$ in the long time limit.

\begin{figure}
\includegraphics[width=0.97\textwidth]{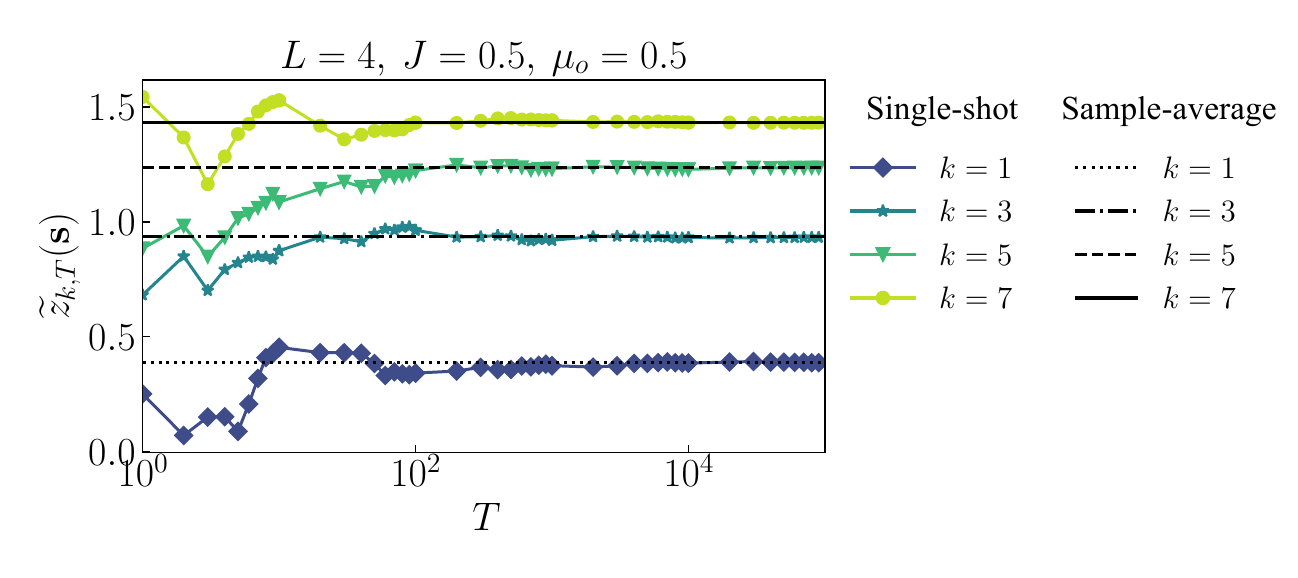}
\caption{Time series of the snapshot Lyapunov spectrum $\widetilde{z}_{k,T}(\bm{s})$ of a single trajectory computed from Eq.~(\ref{eq:SM:snapshot Lyapunov spectrum}) for $L=4$, $J=0.5$, and $\mu_o=0.5$. The boundary condition is the OBC.
Black lines are non-negative Lyapunov spectra computed at $T=10^5$ averaged over different $100$ trajectories for modes with $k=1,3,5,$ and $7$.}
\label{fig:SM:LEevol_L4J5mu50}
\end{figure}

While it requires a special treatment to rigorously show the convergence of the Lyapunov spectrum in monitored settings~\cite{Benoist19} as mentioned in the main text, here we numerically confirm that the Lyapunov spectrum computed for a single trajectory converges to a specific value after sufficiently long time and remains almost unchanged under further time evolution of the circuit, as shown in Fig.~\ref{fig:SM:LEevol_L4J5mu50}.
Additionally, these values closely match the values averaged over $100$ different trajectories.
These results indicate that $z_{\ell, T}(\bm{s})$ becomes independent of the time $T$ and trajectory $\bm{s}$, which allows us to write $z_{\ell, T}(\bm{s})$ as $z_\ell$ for sufficiently large $T$.
In this paper, hence, we only look at a typical single trajectory and consider the temporal average of quantities after the Lyapunov spectrum becomes stationary when we perform Lyapunov analysis, unless otherwise mentioned.
Specifically, we determine the time at which the Lyapunov spectrum is sufficiently stationary{, on the basis of} the following criteria.
First, we evolve the circuit and compute $d\widetilde{z}_{j,t}(\bm{s}) = \widetilde{z}_{2j+1,t}(\bm{s})-\widetilde{z}_{2j-1,t}(\bm{s})$ for all $j=1,\ldots,L-1$ at each time after $t=10^4$.
Then, we compute the averages and the standard deviations of $d\widetilde{z}_{j,t}(\bm{s})$ over the last $1000$ steps, and if the ratios of the standard deviations to the average are less than $\sqrt{10}\times10^{-3}$ for all $j$, we stop the calculation and compute physical quantities averaged over the last $1000$ steps.

\section{Bipartite mutual information and topological entanglement entropy}

\begin{figure}
\includegraphics[width=0.97\textwidth]{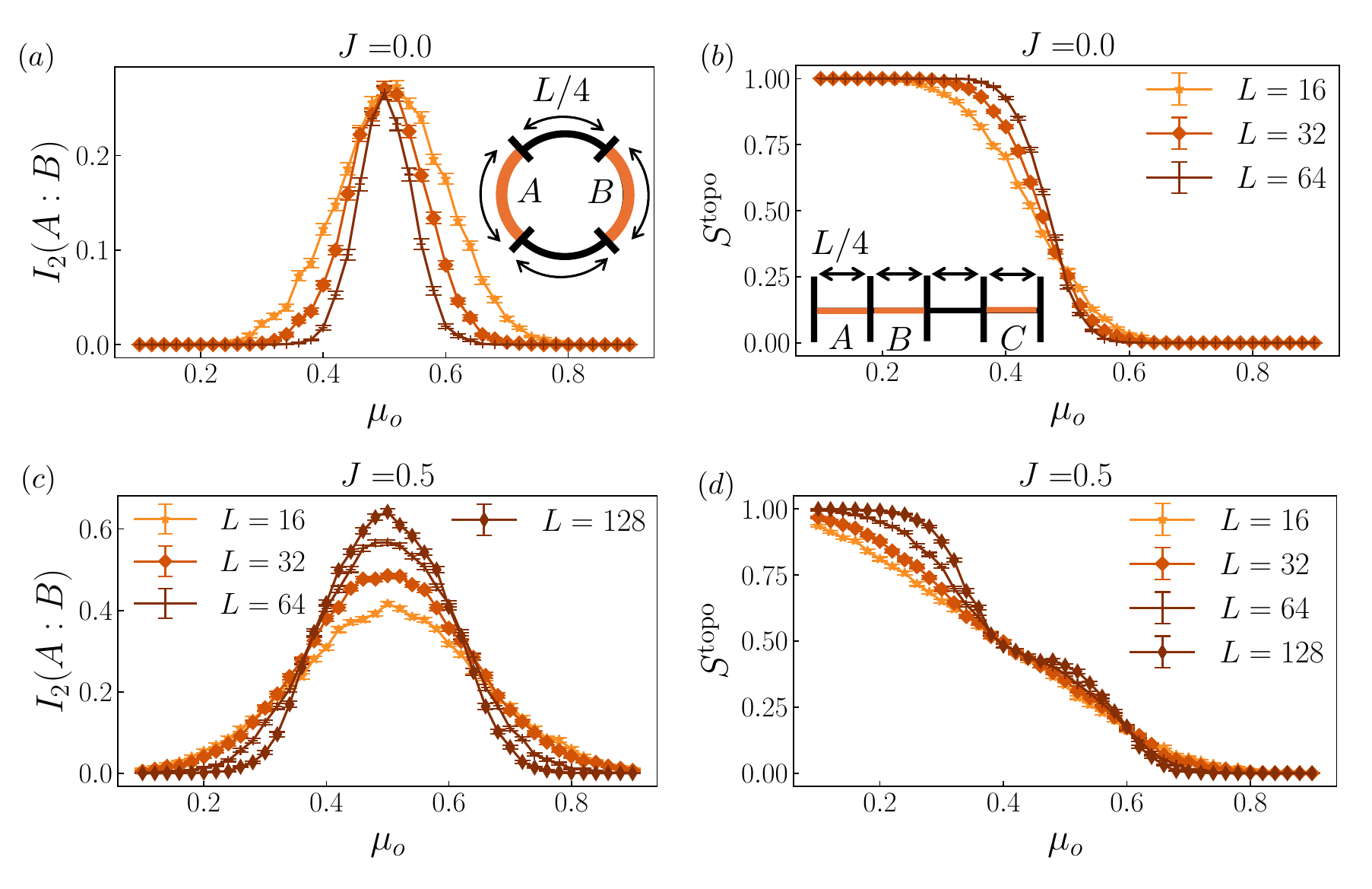}
\caption{(a) Bipartite mutual information and (b) topological entanglement entropy plotted against $\mu_o$ for the measurement-only circuit $(J=0)$.
The bipartite mutual information is computed for the partition of the system into four subsystems of the equal length $L/4$ under the PBC, whereas the topological entanglement entropy is computed for the same partition under the OBC. 
Similar plots are made in (c) and (d) for the monitored circuit with the unitary dynamics $(J=0.5)$.
}
\label{fig:SM:TEEandBMI}
\end{figure}
Our model exhibits a phase transition from the topological area-law phase to the trivial area-law phase as $\mu_o$ increases, when the unitary dynamics is absent.
In the presence of the unitary dynamics, the phase transitions occur twice:
One is the transition from the topological area-law phase to the critical phase, and the other is the transition from the critical phase to the trivial area-law phase.
To confirm this, we first examine the behaviors of conventionally studied quantities, i.e., the bipartite mutual information and topological entanglement entropy (see, e.g., \cite{Lavasani21, Jian23, Pan24}).
The bipartite mutual information between subsystems $A$ and $B$ is given by
\begin{align}
    I_2(A:B) = S_A + S_B - S_{AB}.
\end{align}
We study $I_2(A:B)$ for the circuits with the PBC and consider the partition of the system as shown in Fig.~\ref{fig:SM:TEEandBMI}(a).
Next, the topological entanglement entropy is defined by \cite{Fromholz20}
\begin{align}
    S^{\mathrm{topo}} = S_{AB} + S_{BC} - S_B - S_{ABC},
\end{align}
for the circuits with the OBC and for the partition of the system as shown in Fig.~\ref{fig:SM:TEEandBMI}(b).
Note that $S_X$ in the above expressions represents the von Neumann entanglement entropy of a subsystem $X$, which can be computed from the correlation matrix in the following way.
We first construct a submatrix by extracting only the space corresponding to the {subsystem} $X$ from the original correlation matrix $ (C_{x,x'})|_{x,x'\in\bm{X}}$ with $\bm{X}=[i_1,\ldots,i_{|X|},L+i_1,\ldots,L+i_{|X|}]$.
From its eigenvalues $\{ \lambda_i, 1-\lambda_i \}_{i=1,\ldots,|\bm{X}|/2}$ with $\lambda_i \geq 0$, we can obtain the von Neumann entanglement entropy by
\begin{align}
    S_X = -\sum_{i=1}^{|\bm{X}|/2}(\lambda_i\log_2\lambda_i + (1-\lambda_i)\log_2(1-\lambda_i)).
\end{align}

We now argue that at a phase transition point described by (1+1)-dimensional conformal field theory (CFT), both bipartite mutual information $I_2(A:B)$ and topological entanglement entropy $S^\textrm{topo}$ take constant values independent of the system size, provided that the ratio of the partitions of the system is fixed.
This indicates that the transition point can be estimated from the point at which $I_2(A:B)$ and $S^\textrm{topo}$ computed for different system sizes collapse into a single point.
For the bipartite mutual information $I_2(A:B)$, we consider a one-dimensional chain of the length $L$ with the PBC, which corresponds to an infinite cylinder with the circumference $L$ in the space-time complex coordinate system.
Given the partition $A = [x_1, x_2]$ and $B=[x_3, x_4]$, we need to evaluate the following quantity \cite{Calabrese09b, Chen20},
\begin{align}
\label{eq:CFT}
\frac{\textrm{Tr} \rho_{AB}^n}{\textrm{Tr} \rho_A^n \textrm{Tr} \rho_B^n} = \frac{\langle \mathcal{T}_n(z_1, \bar{z}_1) \mathcal{T}_n(z_2, \bar{z}_2) \mathcal{T}_n(z_3, \bar{z}_3) \mathcal{T}_n(z_4, \bar{z}_4) \rangle}{\langle \mathcal{T}_n(z_1, \bar{z}_1) \mathcal{T}_n(z_2, \bar{z}_2) \rangle \langle \mathcal{T}_n(z_3, \bar{z}_3) \mathcal{T}_n(z_4, \bar{z}_4) \rangle},
\end{align}
where $\rho_X$ is the reduced density matrix for the subsystem $X$, $\mathcal{T}_n(z,\bar{z})$ is the twist field of the conformal dimension $\Delta_n = \bar{\Delta}_n = c(n-n^{-1})/24$, and $z_i = ix_i$ with $x_i \in \mathbb{R}$.
After the conformal mapping $w = e^{2\pi z/L}$ onto the infinite complex plane, we can evaluate multi-point correlation functions of the twist fields in r.h.s of Eq.~\eqref{eq:CFT} from global conformal invariance and find
\begin{align} \label{eq:BMI}
\frac{\textrm{Tr} \rho_{AB}^n}{\textrm{Tr} \rho_A^n \textrm{Tr} \rho_B^n} = F(\eta),
\end{align}
which is a function of the cross ratio $\eta$ defined by
\begin{align}
\eta = \frac{|w_1-w_2||w_3-w_4|}{|w_1-w_3||w_2-w_4|} = \frac{\sin (\pi |x_1-x_2|/L) \sin (\pi|x_3-x_4|/L)}{\sin (\pi|x_1-x_3|/L) \sin(\pi|x_2-x_4|/L)}.
\end{align}
We note that $z_i$ and $w_i$ used here are not related to $z_\ell$ and $\vec{w}_{\ell,T}(\bm{s})$ used for Lyapunov analysis. 
For the partition of the system into four segments of the equal length $L/4$, the cross ratio is independent of the system size $L$ ($\eta=1/2$), and so is the bipartite mutual information $I_2(A:B)$, which can be computed from the limit $n \to 1$ of the logarithm of Eq.~\eqref{eq:BMI}.

For the topological entanglement entropy $S^\textrm{topo}$, we consider a one-dimensional chain of the length $L$ with the OBC, which corresponds to an infinite strip of the width $L$. 
Given the partition $A=[0,x_1]$, $B=[x_1,x_2]$, and $C=[x_3,L]$, we need to evaluate
\begin{align}
\frac{\textrm{Tr} \rho_{AB}^n \textrm{Tr} \rho_{BC}^n}{\textrm{Tr} \rho_B^n \textrm{Tr} \rho_{ABC}^n} = \frac{\langle \mathcal{T}_n(z_2,\bar{z}_2) \rangle \langle \mathcal{T}_n(z_1,\bar{z}_1) \mathcal{T}_n(z_2,\bar{z}_2) \mathcal{T}_n(z_3,\bar{z}_3) \rangle}{\langle \mathcal{T}_n(z_1,\bar{z}_1) \mathcal{T}_n(z_2,\bar{z}_2) \rangle \langle \mathcal{T}_n(z_2,\bar{z}_2) \mathcal{T}_n(z_3,\bar{z}_3) \rangle}.
\end{align}
After the conformal mapping $w'=e^{\pi z/L}$ onto the upper half plane, the multi-point correlation functions of the twist field $\mathcal{T}_n$ in r.h.s can be evaluated in the infinite complex plane via the method of images \cite{CFT}, which yields
\begin{align}
\frac{\textrm{Tr} \rho_{AB}^n \textrm{Tr} \rho_{BC}^n}{\textrm{Tr} \rho_B^n \textrm{Tr} \rho_{ABC}^n} = \frac{\langle \mathcal{T}_n(w'_2) \mathcal{T}_n(\bar{w}'_2) \rangle \langle \mathcal{T}_n(w'_1) \mathcal{T}_n(\bar{w}'_1) \mathcal{T}_n(w'_2) \mathcal{T}_n(\bar{w}'_2) \mathcal{T}_n(w'_3) \mathcal{T}_n(\bar{w}'_3) \rangle}{\langle \mathcal{T}_n(w'_1) \mathcal{T}_n(\bar{w}'_1) \mathcal{T}(w'_2) \mathcal{T}(\bar{w}'_2) \rangle \langle \mathcal{T}_n(w'_2) \mathcal{T}_n(\bar{w}'_2) \mathcal{T}_n(w'_3) \mathcal{T}_n(\bar{w}'_3) \rangle}.
\end{align}
Since this involves a six-point function of $\mathcal{T}_n$, global conformal invariance dictates that it should be a function of three cross ratios constructed out of $w'_j, \bar{w}'_j \ (j=1,2,3)$ \cite{Calabrese09b}, 
\begin{align} \label{eq:TEE}
\frac{\textrm{Tr} \rho_{AB}^n \textrm{Tr} \rho_{BC}^n}{\textrm{Tr} \rho_B^n \textrm{Tr} \rho_{ABC}^n} = F'(\eta'_1, \eta'_2, \eta'_3),
\end{align}
where
\begin{align}
\eta'_1 &= \frac{(w'_1-\bar{w}'_1)(w'_2-\bar{w}'_2)}{(w'_1-w'_2)(\bar{w}'_1-\bar{w}'_2)} = \frac{\sin (\pi x_1/L) \sin (\pi x_2/L)}{\sin^2 [\pi(x_1-x_2)/2L]}, \\
\eta'_2 &= \frac{(w'_2-\bar{w}'_2)(w'_3-\bar{w}'_3)}{(w'_2-w'_3)(\bar{w}'_2-\bar{w}'_3)} = \frac{\sin (\pi x_2/L) \sin (\pi x_3/L)}{\sin^2 [\pi(x_2-x_3)/2L]}, \\
\eta'_3 &= \frac{(w'_1-\bar{w}'_1)(w'_3-\bar{w}'_3)}{(w'_1-w'_3)(\bar{w}'_1-\bar{w}'_3)} = \frac{\sin (\pi x_1/L) \sin (\pi x_3/L)}{\sin^2 [\pi (x_1-x_3)/2L]}.
\end{align}
For the partition of the system into four segments of the equal length $L/4$, the cross ratios become $\eta'_1=\eta'_2=2/(\sqrt{2}-1)$ and $\eta'_3=1$. 
Since they are independent of the system size $L$, the topological entanglement entropy obtained by the logarithm of Eq.~\eqref{eq:TEE} is also independent of $L$.

Now, let us investigate our monitored circuits.
We show that the behaviors of the bipartite mutual information and topological entanglement entropy for our circuits are consistent with those for the CFT described above.
In Fig.~\ref{fig:SM:TEEandBMI}(a) and (b), we plot the bipartite mutual information and topological entanglement entropy for the measurement-only circuit with $J=0$ for $L=16, 32, $ and $64$, respectively.
The bipartite mutual information shows a broad peak in the vicinity of $\mu_o=0.5$, and the values of the curves of different system sizes at $\mu_o=0.5$ are close.
The topological entanglement entropy shows a clear crossing in the vicinity of $\mu_o=0.5$.
These results imply that the topological phase transition occurs at $\mu_o \simeq 0.5$.

In Fig.~\ref{fig:SM:TEEandBMI}(c) and (d), we plot $I_2(A:B)$ and $S^{\mathrm{topo}}$ for the circuit with unitary dynamics $J=0.5$ for $L=16, 32, 64, $ and $128$, respectively.
Both quantities appear to show two scale-invariant points at $\mu_o \simeq 0.4$ and $0.6$, indicating the entanglement transitions between the topological/trivial area-law phase and the critical phase.
While the crossing points of $I_2(A:B)$ drift as the system size increases due to the large finite size effect, different curves of $S^{\mathrm{topo}}$ clearly collapse to a single point at $\mu_o \simeq 0.4$ and $0.6$.
From these data, we have roughly determined the location of the entanglement transitions as $\mu_o \simeq 0.4$ and $0.6$.
Note that Ref.~\cite{Fava23} argued that the phase transitions between the area-law phases and critical phase for weakly monitored Majorana fermions are described by a scale-invariant theory, which is also consistent with our results.

\section{Lyapunov gap and dynamics of topological invariant}
\begin{figure}
\includegraphics[width=0.97\textwidth]{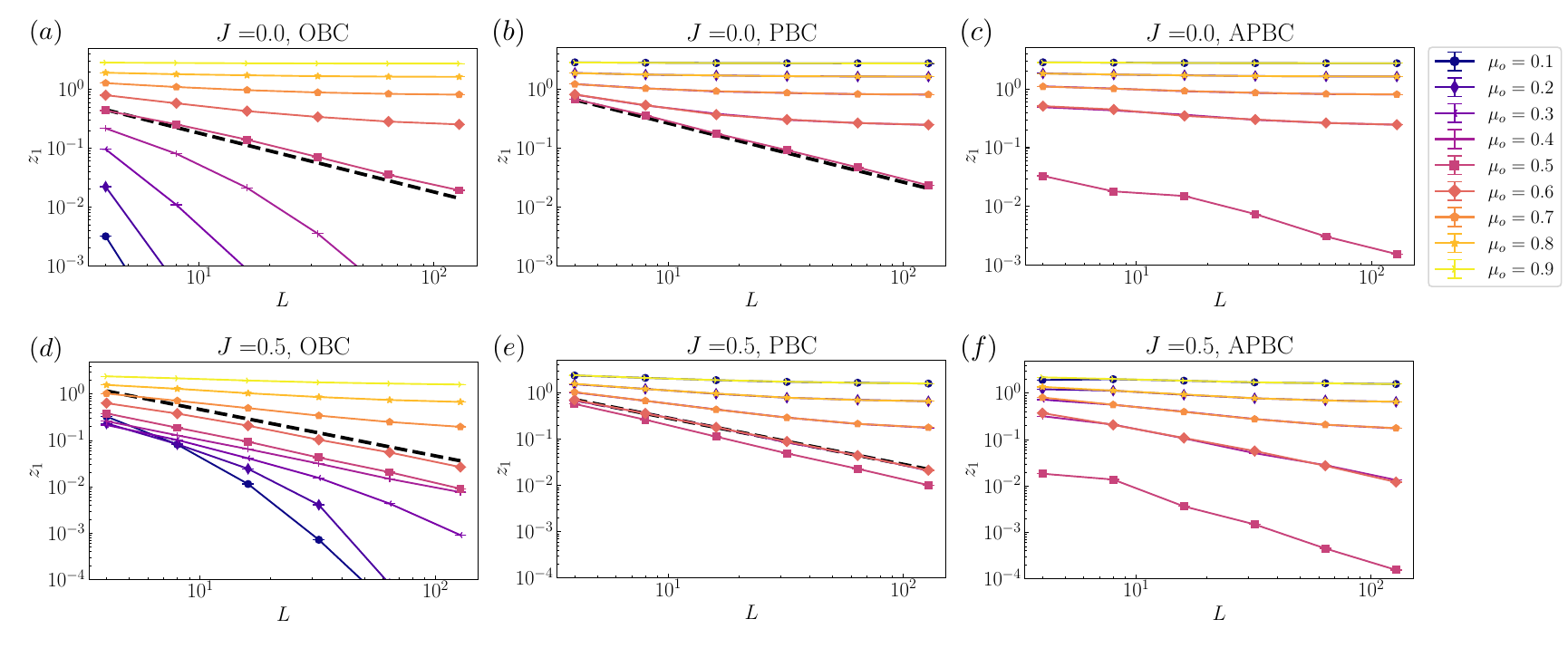}
\caption{The lowest non-negative Lyapunov spectrum with respect to $L$ for the circuit with $J=0$ under the (a) OBC, (b) PBC, and (c) APBC; and with $J=0.5$ under the (d) OBC, (e) PBC, and (f) APBC.
Each black dashed line in panels (a,b,d,e) represents a trend proportional to $1/L$.
}
\label{fig:SM:z1scaling}
\end{figure}
In this section, we first study the scaling form of the lowest non-negative single-particle Lyapunov spectrum $z_1$ with respect to system sizes $L$. The  $z_1$ is particularly important because it corresponds to the lowest energy gap of the many-body effective Hamiltonian $\hat{\mathcal{H}}_{\mathrm{eff},T}(\bm{s})$ for large $T$.

In Fig.~\ref{fig:SM:z1scaling}(a,d), we show $z_1$ against $L=4,8,16,32,64$, and $128$ in the monitored circuit under the OBC with (a) $J=0$ and (d) $J=0.5$.
In the measurement-only case ($J=0$), $z_1$ shows a decay faster than the power law in the topological phase ($\mu_o \lesssim 0.5)$, while it approaches a finite value in the trivial area-law phase.
These results are similar in the presence of unitary dynamics ($J=0.5$) as well, while $z_1$ appears to decay slowly even in the trivial area-law phase for our available system sizes.
At the topological transition ($\mu_o\simeq0.5)$ in the measurement-only circuit, $z_1$ shows a power-law decay with its exponent close to $-1$.
We can see a similar behavior in the circuit with unitary evolution, while we cannot conclude the true scaling form especially inside the critical phase ($0.4\lesssim\mu_o\lesssim0.6)$.

Next, the spectrum $z_1$ in the monitored circuit under the PBC with $J=0$ and $0.5$ are plotted in Fig.~\ref{fig:SM:z1scaling}(b) and (e), respectively.
In the main text, we have discussed the relaxation time of the topological invariant $\tau_{\mathrm{relax}}$.
As $z_1$ corresponds to the many-body gap of the effective Hamiltonian, it characterizes the timescale of the relaxation, i.e.,  $\tau_{\mathrm{relax}}\gtrsim {1/z_1}$.
Here, we can see that $z_1$ scales as $1/L$ near the phase boundaries, while it shows a decay slightly faster than $1/L$ inside the critical phase, leading to $\tau_{\mathrm{relax}}\sim\mathcal{O}(L)$ at the transition points and $\tau_{\mathrm{relax}}>\mathcal{O}(L)$ inside the critical phase.
On the other hand, as $z_1$ decays slower than $1/L$ (or does not decay with respect to $L$) in the topological and trivial area-law phases, we can conclude $\tau_{\mathrm{relax}}<\mathcal{O}(L)$.
These features can also be seen in the monitored circuit under the APBC except for $\mu_o = 0.5$, as shown in Fig.~\ref{fig:SM:z1scaling}(c,f).

\begin{figure}
\includegraphics[width=0.97\textwidth]{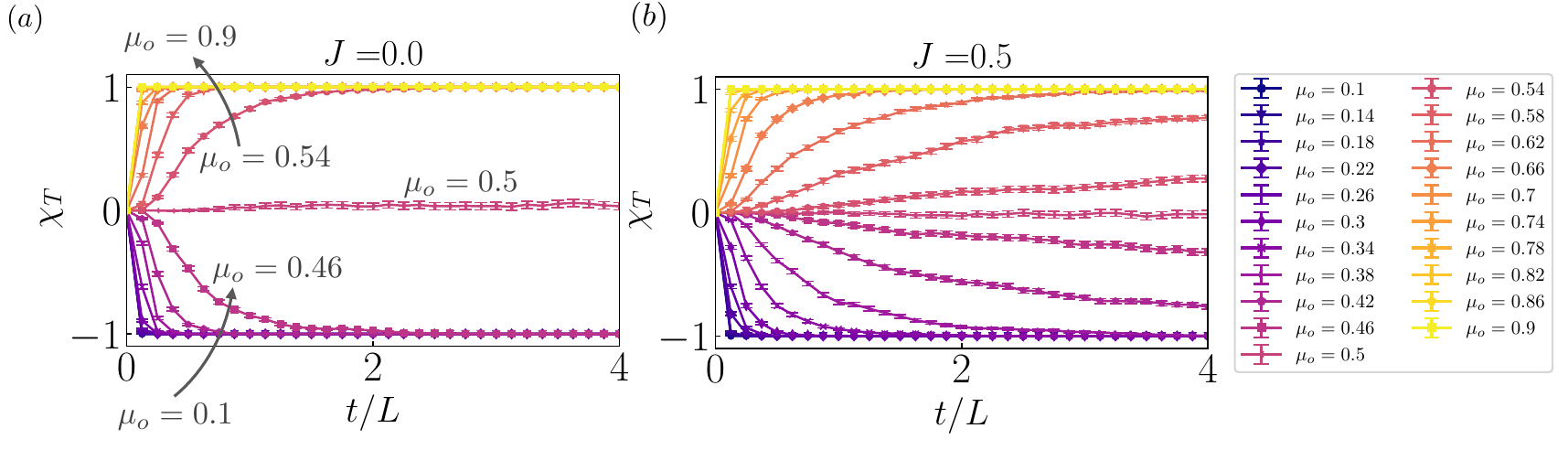}
\caption{Time series of the topological invariant $\chi_T$ for $L=64$ with (a) $J=0$ and (b) $J=0.5$ averaged over $1000$ trajectories.}
\label{fig:SM:ChiEvol}
\end{figure}
The above argument on the timescale is numerically confirmed by the time series of the topological invariant $\chi_T$ in the circuits with $J=0$ and $0.5$ as shown in Fig.~\ref{fig:SM:ChiEvol}(a) and (b), respectively.
In the measurement-only circuit, as $z_1$ for the PBC and APBC become finite except for $\mu_o \simeq 0.5$, $\chi_T$ rapidly converges to $\pm1$ for $T=\mathcal{O}(L)$.
On the other hand, in the circuit with $J=0.5$, $\chi_T$ rapidly converges to $\pm1$ for $T=\mathcal{O}(L)$ in the area-law phases with $\mu_o\lesssim0.4$ or $\mu_o\gtrsim0.6$, where $z_1$ shows a decay slower than $1/L$ (or no decay), while it takes much longer time to converge inside the critical phase.

Finally, let us discuss $\mu_o=0.5$. For both cases with $J=0$ and $0.5$, Fig.~\ref{fig:SM:z1scaling}(c,f) shows that $z_1$ at $\mu_o=0.5$ is about ten times smaller than those at the other $\mu_o$.
This implies that there exists an exact gap closing in the APBC circuit near $\mu_o=0.5$.
Because of the small $z_1$, $\tau_\mathrm{relax}$ becomes much longer at this point.
This leads to the fact that $\chi_T$ with $\mu_o=0.5$ takes a value close to zero for much longer times than those with the other values of $\mu_o$, as shown in Fig.~\ref{fig:SM:ChiEvol}.


\end{document}